# EFFECTIVE PIEZOELECTRIC RESPONSE OF TWIN WALLS IN FERROELECTRICS

Anna N. Morozovska,[1,*] Eugene A. Eliseev,[2] Olexander V. Varenyk,[3] and Sergei V. Kalinin[4,†]

[1] Institute of Physics, National Academy of Sciences of Ukraine,
41, pr. Nauki, 03028 Kiev, Ukraine

[2] Institute for Problems of Materials Science, National Academy of Sciences of Ukraine,
3, Krjijanovskogo, 03142 Kiev, Ukraine

[3] Taras Shevchenko Kiev National University, Radiophysical Faculty
4g, pr. Akademika Hlushkova, 03022 Kiev, Ukraine

[4] The Center for Nanophase Materials Sciences, Oak Ridge National Laboratory,
Oak Ridge, TN 37831

**Abstract**

The effective piezoelectric coefficients of twin walls in tetragonal ferroelectric are calculated in the framework of decoupling approximation and Landau-Ginzburg-Devonshire theory allowing for polarization gradient terms, electrostriction and flexoelectric coupling. Using an example of piezoelectric response of $a_1$-$a_2$ twins to a homogeneous electric field, we show that the response is almost independent on the flexoelectric coupling, but is very sensitive to the values of polarization gradient coefficients. This behavior originates from the strong coupling between local dielectric susceptibility and the gradient coefficients. The enhancement of piezoelectric response from 10% up to a factor of $10^3$ times is predicted. The local electromechanical response of the domain walls can thus provide information on the gradient terms in GLD expansion and pinning mechanisms of the ferroelectric domain walls. The observability of these effects by the piezoresponse force microscopy of electroded structures and impact on the functional properties of the systems with dense domain structures is analyzed.

[*] morozo@i.com.ua

[†] sergei2@ornl.gov

## 1. INTRODUCTION

Domain structures and domain walls have long been recognized as the key factor determining the functionality of ferroelectric materials, directly controlling behaviors such as polarization and electromechanical hysteresis, dielectric and piezoelectric nonlinearities, and many others. Recently, functionality of ferroelectric domain walls per se have emerged as a new paradigm in design of ferroelectric and multiferroic materials and structures, with the examples ranging from static and dynamic domain wall conduction, enhanced electromechanical responses, magnetoelectric coupling, and many others [1, 2, 3, 4, 5]. Much of this progress was stimulated by the recent emergence of scanning probe microscopy techniques such as Piezoresponse Force Microscopy and conductive atomic force microscope that enabled local probing of domain wall properties. Modern advances in PFM of ferroelectric domains and nanostructures are reviewed by Gruverman and Kholkin [6], Kalinin *et al* [7] and Gruverman [8]. These developments, in turn, necessitate the theoretical analysis of electromechanical and conduction phenomena on a single wall level, with the dual goal of interpretation of PFM contrast and exploring emergent domain wall functionalities.

The necessity for calculating the Piezoresponse Force Microscopy (PFM) signal for materials of general symmetry, as well as calculation of response at micro- and nanostructural elements such as periodic domains, isolated cylindrical domains, and topographically inhomogeneous ferroelectrics such as nanoparticles have stimulated theoretical attempts to derive approximate solutions for position-dependent PFM signal in inhomogeneous materials [9, 10, 11, 12, 13, 14] A general approach for the calculation of the electromechanical response is based on the decoupling approximation typically using for the piezoelectric response calculations of domain structures with "frozen" polarization, i.e. the polarization does not change during imaging process (see e.g. [6-8]). A simplified 1D version of the decoupling model was originally suggested by Ganpule [15] to account for the effect of 90° domain walls on PFM imaging. A similar 1D approach was adapted by Agronin *et al* [16] to yield closed-form solutions for the PFM signal. The 3D version of this approach was developed by Felten *et al* [17] using the analytical form for the corresponding Green's function. Independently, Scrymgeour and Gopalan [18] have used the finite element method to model PFM signals across the 180-degree domain walls. Kalinin *et al* [19], Eliseev *et al* [20] and Morozovska *et al* [21] have applied the decoupling theory to derive analytical expressions for the PFM response of materials with low symmetry, the PFM resolution function and 180-degree domain wall profiles, and interpret PFM spectroscopy data. Kalinin et al. [22] found the analytical expression for lateral resolution of PFM for 180-degree domains in capacitor structures. Lei et al. [23] analyzed different mechanisms of lateral and vertical PFM response formation for the case of 3m ferroelectrics. Using decoupling approximation, Pan et al. [24] developed a numerical integration scheme to analyze the expected PFM response in ferroelectrics with arbitrary domain configurations, including the case of the depth distribution of piezoelectric coefficients. Recently they applied their approach to estimation of the lateral resolution of twin domain walls PFM response as well as to the consideration of depth resolution [25].

Note, that most of the authors [9-25] analyzed the PFM contrast in the rigid polarization approximation, namely they considered PFM response frozen and regard ferroelectric polarization independent on the tip voltage during the imaging process. Only recently Yang and Dayal [26, 27] performed "soft" 3D phase-field simulations to model PFM response of various domain structures, including 90 degree domains for different boundary conditions and the mechanisms of electric field screening.

As it follows from the brief overview of current state-of-art in PFM, the local piezoelectric response of ferroelectric twin walls requires analytical theoretical studies, since only numerical results are available [25], in contrast to the relatively well-studied response of 180-degree walls. This interest is strongly enhanced by recent theoretical studies and experimental observations. For example, Sluka *et al* [28] showed that ferroelectrics like $BaTiO_3$ with dense patterns of 90-degree charged domain walls are expected to have strongly enhanced piezoelectric properties, that can open the way for new functionalities. Karthik et al. [29] reported about the contribution of 90° ferroelastic domain walls to the dielectric permittivity in thin films $PbZr_{0.2}Ti_{0.8}O_3$. They experimentally observed a strong enhancement of the permittivity with increasing domain wall density that matches the predictions of the phenomenological GLD models. This motivates us to **calculate analytically** the effective piezoelectric coefficients of twin walls in tetragonal ferroelectric in the framework of decoupling approximation and LGD theory.

The paper is organized as following. Mathematical statement of the general problem of domain wall response for dynamic polarization is given in the Section 2. Section 3 contains the derivation of the problem analytical solution in decoupling approximation and the analyses for different strength of polarization gradient and flexoelectric coupling. Section 4 is a brief summary.

## 2. STATEMENT OF THE PROBLEM

Here, we consider classical 2-4-6 Landau potential [30, 31] including polarization gradient terms, electrostriction and flexoelectric coupling. Equilibrium polarization distribution is then given by the Euler-Lagrange equations [32] obtained from the Landau potential minimization with respect to polarization components:

$$\left(\alpha\delta_{ij} + 2u_{mn}q_{mnij}\right)P_j + \alpha_{ijkl}P_jP_kP_l + \alpha_{ijklmn}P_jP_kP_lP_mP_n - g_{ijkl}\frac{\partial^2 P_k}{\partial x_j \partial x_l} = f_{mnli}\frac{\partial u_{mn}}{\partial x_l} + E_i \qquad (1)$$

$\delta_{ij}$ represents the Kroneker symbol. Hereinafter $u_{kl}(\mathbf{r})$ denotes the elastic strain, $P_m(\mathbf{r})$ denotes electric polarization, $q_{ijkl}$ denotes the electrostriction stress tensor, and $f_{ijkm}$ the flexoelectric tensor. Film-substrate misfit strains can be accounted through the renormalization of the free energy coefficients $\alpha(T)$ and $\alpha_{ijkl}$ [33]. The electric field is $E_k(\mathbf{r}) = -\partial \varphi(\mathbf{r})/\partial x_k$, $\varphi(\mathbf{r})$ is the electric potential.

Allowing for the gradient term contribution, the natural boundary conditions for Eq.(2a) are obtained after the minimization of free energy:

$$g_{ijkl} n_j \left.\frac{\partial P_k}{\partial x_l}\right|_S = 0 \qquad (2)$$

$S$ is the ferroelectric surface. Note that this condition is consistent with the quasi-homogeneous distribution of polarization, since we do not consider the surface energy contribution to free energy, which leads to the intrinsic distribution of spontaneous polarization in nanosized ferroelectrics [34].

Under the absence of the screening charges, the Poisson-type equation for electrostatic potential $\varphi(\mathbf{r})$ is:

$$\varepsilon_0 \varepsilon_b \frac{\partial^2 \varphi}{\partial x_i \partial x_i} = \frac{\partial P_k}{\partial x_k}. \qquad (3)$$

Hereinafter $\varepsilon_b$ is background permittivity and $\varepsilon_0 = 8.85 \times 10^{-12}$ F/m is the universal dielectric constant. Boundary conditions to Eq.(2) have the form:

$$\varphi(x_1, x_2, 0) = V_0(x_1, x_2), \qquad \varphi(x_1, x_2, h) = 0 \qquad (4)$$

where $V_0(x_1, x_2)$ denotes the electrostatic potential distribution at the sample surface $x_3 = 0$ including (if necessary) the constant built-in potential [35, 36], which originate from surface dipole layers, e.g. due to the Shottky barrier at the tip (or electrode) – surface junction, $h$ is the film thickness.

Elastic equations of state follow from the variation of the free energy on the strain tensor, $\sigma_{ij} = c_{ijkl} u_{kl} + f_{ijkl} \frac{\partial P_l}{\partial x_k} + q_{ijkl} P_k P_l$, where $\sigma_{ij}$ is the stress tensor. As the strain tensor is given by $u_{ij} = \frac{1}{2}\left(\frac{\partial u_i}{\partial x_j} + \frac{\partial u_j}{\partial x_i}\right)$, the Lame-type equation for mechanical displacement $u_i$ can be obtained from the equation of mechanical equilibrium $\partial \sigma_{ij}/\partial x_i = 0$, where the stress tensor $\sigma_{ij}(\mathbf{r})$ is

$$c_{ijkl} \frac{\partial^2 u_k}{\partial x_j \partial x_l} = -\frac{\partial}{\partial x_j}\left(f_{ijkl} \frac{\partial P_l}{\partial x_k} + q_{ijkl} P_k P_l\right) \qquad (5)$$

Mechanical boundary conditions [37] corresponding to typical experiments are defined on a mechanically free interface, $z = 0$, where the normal stress is absent (more specifically, tip-surface or electrode-surface forces are small),

$$\sigma_{3i}(x_1, x_2, 0) = 0, \tag{6a}$$

and on substrate interface $x_3 = h$, where the displacement $u_i$ is zero and in-plane misfit strains are fixed for a thick "rigid" cubic substrate

$$u_i(x_1, x_2, h) = 0, \quad u_{11}(x_1, x_2, h) = u_{22}(x_1, x_2, h) = u_m. \tag{6b}$$

or continuous for a "soft matched" thin substrate or becomes unnecessary for a semi-infinite sample.

Note that Eqs.(1), (3) and (5) are coupled, because polarization depends on the strain and external field. Hence, the formal solution of Eq.(5) written via elastic Green function as

$$u_i = -\iiint\limits_{0<\xi_3<h} \frac{\partial G_{ij}^S(x_1-\xi_1, x_2-\xi_2, x_3, \xi_3)}{\partial \xi_m} \left( f_{mjkl} \frac{\partial P_l}{\partial \xi_k} + q_{mjkl} P_k P_l \right) d^3\xi \tag{7}$$

is rigorously defined only in decoupling approximation.

### 3. DECOUPLING APPROXIMATION

In decoupling approximation [15, 17, 18, 19, 20, 21] we regard that elastic field in Eq.(3a) is the known function of polarization. Here, we analyze the structure of the elastic fields created by the twin domain wall – surface junction using the perturbation method proposed by Rychetsky [38]. In the first approximation, the strains $u_{ij}^S(\mathbf{x})$ at location $\mathbf{x}$ from the surface induced by the elastic wall – surface junction is given by the convolution of the corresponding Green function with the elastic stress field, $\sigma_{jk}^0$, unperturbed by the surface influence:

$$u_{ij}^S(\mathbf{x}) = u_{ij}^0(\mathbf{x}) + \int_{-\infty}^{\infty} d\xi_1 \int_{-\infty}^{\infty} d\xi_2 \left( \frac{\partial G_{ik}(\mathbf{x}-\xi)}{2\partial x_j} + \frac{\partial G_{jk}(\mathbf{x}-\xi)}{2\partial x_i} \right) \sigma_{km}^0(\xi_1, \xi_2) n_m \tag{8}$$

Corresponding Green's tensor $G_{ij}(\mathbf{x}-\xi) \equiv G_{ij}(x_1-\xi_1, x_2-\xi_2, x_3)$ for elastically isotropic half-space is given by Lur'e [39] and Landau and Lifshitz [40] (see **Appendix A**); $n_k$ is the outer normal to the mechanically free surface $x_3 = 0$. Here we consider the semi-infinite mechanically free crystal, but not the film on the substrate. However, the approach can be extended to the film case if one will use the Green function corresponding to the elastic problem of mechanically clamped/free film. Below we consider a matched substrate or a semi-infinite sample for the sake of elastic problem solution simplicity.

#### 3.1. Basic equations for the twin-wall surface – junction

Here we consider the 90-degree $a1$-$a2$ twin-wall surface - junction in a tetragonal ferroelectric (e.g. BaTiO$_3$, PbTiO$_3$ at room temperature) as shown in **Figure 1a**.

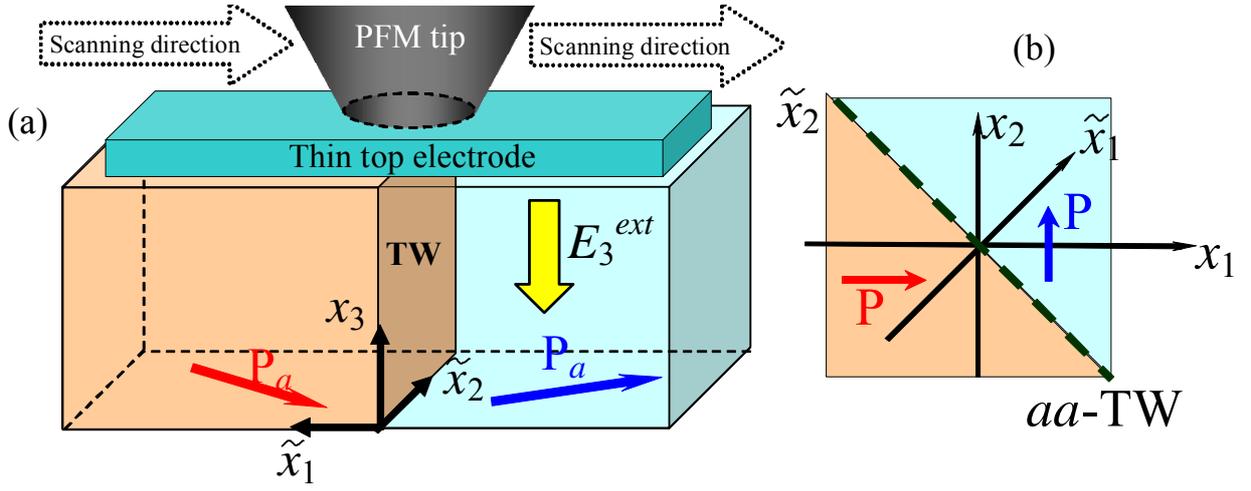

**FIGURE. 1**. **Probing geometry of *a*1-*a*2 twin domain –surface junctions (a). (b)** Spontaneous polarization vector is $\mathbf{P}=(P_S,0,0)$ in a1-domain and $\mathbf{P}=(0,-P_S,0)$ in a2-domain. Note, that under the condition of nonzero polarization gradient inherent to the adopted LGD model, ac-twins appearance can be suppressed by depolarization effects.

Below we introduced 45°-rotated coordinate system (see e.g. Ref. [41]) with one axis, $\tilde{x}_3 \equiv x_3$, and two other rotated axes, $\tilde{x}_1 = (x_1+x_2)/\sqrt{2}$ perpendicular to the domain wall, and $\tilde{x}_2 = (-x_1+x_2)/\sqrt{2}$ parallel to the domain wall (see **Fig. 1b**). For the considered geometry of the problem in the rotated system the distributions of polarization are $\tilde{P}_1 = (P_1+P_2)/\sqrt{2}$ and $\tilde{P}_2 = (P_2-P_1)/\sqrt{2}$. Very far from the TW boundary $\tilde{x}_1 = 0$ polarization components $\tilde{P}_1(\tilde{x}_1 \to \pm\infty) = P_S/\sqrt{2}$ and $\tilde{P}_2(\tilde{x}_1 \to \pm\infty) = \mp P_S/\sqrt{2}$. Polarization component $\tilde{P}_3 \equiv P_3$ is absent under the condition $E_3^{ext}=0$.

Inhomogeneous elastic stresses $\sigma_{jk}^0$ originate from the electrostriction and flexoelectric coupling appeared in the vicinity of the twins. For the considered geometry and small external fields the inequality $|\tilde{P}_3| \ll |\tilde{P}_{1,2}|$ is valid and unperturbed stresses and strains have the form:

$$\tilde{\sigma}_{22}^0(\tilde{x}_1) = \frac{s_{11}U_2 - s_{12}U_3}{s_{11}\tilde{s}_{11} - s_{12}^2}, \quad \tilde{\sigma}_{33}^0(\tilde{x}_1) = \frac{\tilde{s}_{11}U_3 - s_{12}U_2}{s_{11}\tilde{s}_{11} - s_{12}^2}, \quad \tilde{\sigma}_{23}^0 = \frac{U_4}{s_{44}}, \quad \tilde{\sigma}_{11}^0 = \tilde{\sigma}_{13}^0 = \tilde{\sigma}_{12}^0 = 0, \quad (9a)$$

$$\tilde{u}_{11}^0 = -\tilde{F}_{11}\frac{\partial \tilde{P}_1}{\partial \tilde{x}_1} + \tilde{Q}_{11}\tilde{P}_1^2 + \tilde{Q}_{12}\tilde{P}_2^2 + \left(\frac{s_{11}\tilde{s}_{12} - s_{12}^2}{s_{11}\tilde{s}_{11} - s_{12}^2}U_2 + s_{12}\frac{\tilde{s}_{11} - \tilde{s}_{12}}{s_{11}\tilde{s}_{11} - s_{12}^2}U_3\right) \quad (9b)$$

$$\tilde{u}_{22}^0 = (\tilde{Q}_{12} + \tilde{Q}_{11})\frac{P_S^2}{2} + Q_{12}P_3^2, \qquad \tilde{u}_{33}^0 = Q_{12}P_S^2 + Q_{11}P_3^2, \quad (9c)$$

$$\tilde{u}_{23}^0 = \frac{Q_{44}}{2}\tilde{P}_2\,P_3, \qquad \tilde{u}_{13}^0 = \frac{Q_{44}}{2}\tilde{P}_1\,P_3, \qquad \tilde{u}_{12}^0 = -\tilde{F}_{66}\frac{\partial \tilde{P}_2}{\partial \tilde{x}_1} + \tilde{Q}_{66}\tilde{P}_1\tilde{P}_2. \quad (9d)$$

Functions:

$$U_2(\tilde{x}_1) = \tilde{F}_{12}\frac{\partial \tilde{P}_1}{\partial \tilde{x}_1} + \tilde{Q}_{12}\left(\frac{P_S^2}{2} - \tilde{P}_1^2\right) + \tilde{Q}_{11}\left(\frac{P_S^2}{2} - \tilde{P}_2^2\right) - Q_{12}P_3^2, \tag{10a}$$

$$U_3(\tilde{x}_1) = F_{12}\frac{\partial \tilde{P}_1}{\partial \tilde{x}_1} + Q_{12}\left(P_S^2 - \tilde{P}_2^2 - \tilde{P}_1^2\right) - Q_{11}P_3^2, \qquad U_4(\tilde{x}_1) = -Q_{44}\tilde{P}_2 P_3 \tag{10b}$$

Elastic compliances $s_{ijkl}$, electrostriction strain coefficients $Q_{ijkl}$ and flexoelectric tensor $F_{ijkl}$ are written in Voight notations.

Note that elastic fields given by Eq.(9) is consistent with mechanical compatibility relations only for the case $P_3 \equiv 0$. Otherwise the twin wall plane should rotate in the absence of pinning centers. The pinning will clamp the wall and the expressions (9) can be a reasonable approximation for very small external fields in $x_3$-direction.

Euler-Lagrange equations for polarization components $\tilde{P}_i$ depending only on the distance $\tilde{x}_1$ from the wall have the form [42]:

$$\alpha_1\tilde{P}_1 + \beta_1\tilde{P}_1^3 + 6\tilde{a}_{111}\tilde{P}_1^5 - \tilde{g}_{11}\frac{\partial^2 \tilde{P}_1}{\partial \tilde{x}_1^2} = \tilde{E}_1 + \tilde{F}_{12}\frac{\partial \tilde{\sigma}_{22}^0}{\partial \tilde{x}_1} + F_{12}\frac{\partial \tilde{\sigma}_{33}^0}{\partial \tilde{x}_1}, \tag{11a}$$

$$\alpha_2\tilde{P}_2 + \beta_2\tilde{P}_2^3 + 6\tilde{a}_{111}\tilde{P}_2^5 - \tilde{g}_{66}\frac{\partial^2 \tilde{P}_2}{\partial \tilde{x}_1^2} - Q_{44}\tilde{\sigma}_{23}^0 P_3 = \tilde{E}_2. \tag{11b}$$

$$\alpha_3 P_3 + \beta_3 P_3^3 + 6\tilde{a}_{111}P_3^5 - g_{44}\frac{\partial^2 P_3}{\partial \tilde{x}_1^2} - Q_{44}\tilde{\sigma}_{23}^0 \tilde{P}_2 = E_3. \tag{11c}$$

Functions $\alpha_1 = \dfrac{1}{\varepsilon_0\varepsilon_b} + 2\left(a_1 + \tilde{a}_{12}\tilde{P}_2^2 + a_{12}P_3^2 + \tilde{a}_{112}\tilde{P}_2^4 - \tilde{Q}_{12}\tilde{\sigma}_{22}^0 - Q_{12}\tilde{\sigma}_{33}^0\right)$, $\beta_1 = 4\tilde{a}_{11} + 4\tilde{a}_{112}\tilde{P}_2^2$,

$\alpha_2 = 2\left(a_1 + \tilde{a}_{12}\tilde{P}_1^2 + a_{12}P_3^2 + \tilde{a}_{112}\tilde{P}_1^4 - \tilde{Q}_{11}\tilde{\sigma}_{22}^0 - Q_{12}\tilde{\sigma}_{33}^0\right)$, $\beta_2 = 4\tilde{a}_{11} + 4\tilde{a}_{112}\tilde{P}_1^2$,

$\alpha_3 = 2\left(a_1 + a_{12}\left(\tilde{P}_1^2 + \tilde{P}_2^2\right) + \tilde{a}_{113}\left(\tilde{P}_1^4 + \tilde{P}_2^4\right) + \tilde{a}_{123}\tilde{P}_1^2\tilde{P}_2^2 - Q_{12}\tilde{\sigma}_{22}^0 - Q_{11}\tilde{\sigma}_{33}^0\right)$, $\beta_3 = 4a_{11} + 4\tilde{a}_{112}\left(\tilde{P}_2^2 + \tilde{P}_1^2\right)$.

Coefficients $\tilde{a}_{11} = \dfrac{2a_{11} + a_{12}}{4}$, $\tilde{a}_{12} = \dfrac{6a_{11} - a_{12}}{2}$, $\tilde{a}_{111} = \dfrac{a_{111} + a_{112}}{4}$, $\tilde{a}_{112} = \dfrac{15a_{111} - a_{112}}{4}$,

$\tilde{g}_{66} = \dfrac{g_{11} - g_{12}}{2}$ and $\tilde{g}_{11} = \dfrac{g_{11} + g_{12} + 2g_{44}}{2}$. Factor $\dfrac{1}{\varepsilon_0\varepsilon_b}$ in $\alpha_1$ originated from the bare depolarization field $\tilde{E}_1^d = -\tilde{P}_1/\varepsilon_0\varepsilon_b$. In Equations (11a,b) we neglect the nonlinear terms $\tilde{P}_3^2$ since regarded that it is much smaller than $\tilde{P}_{1,2}$.

Electric field $\tilde{E}_i = \tilde{E}_i^{ext} + \tilde{E}_i^{pin}$ is the sum of external field, $\tilde{E}_i^{ext}$, and pinning one, $\tilde{E}_i^{pin}$. Note, that the amplitude of lattice pinning field can be estimated within the framework of Ishibashi model [43, 44, 45] as:

$$E^{pin}(\tilde{x}_1, w) = -2\alpha P_S \left(\frac{w}{a}\right)^3 \exp\left(-\pi^2 \frac{w}{a}\right) e^4 \left(\frac{\pi}{2}\right)^{7/2} \cos\left(\frac{2\pi \tilde{x}_1}{a}\right) \quad (12)$$

For domain wall thickness $w$ more than the lattice constant $a$, this field is at least about 100 times smaller than the thermodynamic coercive field $E_C = -2\alpha P_S/\sqrt{27}$. For example, $E^{pin}(0,a) \approx 0.0274 E_C$, $E^{pin}(0,2a) \approx 1.4 \times 10^{-5} E_C$ and $E^{pin}(0,3a) \approx 2 \times 10^{-10} E_C$. Despite the lattice pinning field is much smaller than the thermodynamic coercive field, it does prevent the wall movement in the infinitely small external field and thus defines the boundary for frozen/unfrozen case. Note that there are pinning due to the impurities adsorbed by the wall to compensate the strain and charge, and these pinning centers are stationary, i.e. they have slow time dynamics.

Equations (11) should be solved numerically along with the boundary conditions very far from the TW:

$$\tilde{P}_1(\tilde{x}_1 \to \pm\infty) = P_S/\sqrt{2}, \quad \tilde{P}_2(\tilde{x}_1 \to \pm\infty) = \mp P_S/\sqrt{2}, \quad P_3(\tilde{x}_1 \to \pm\infty) \propto E_3^{ext} \quad (13)$$

Twin wall (TW) boundary is $\tilde{x}_1 = 0$ and $P_S^2 = \dfrac{-a_1}{\sqrt{a_{11}^2 - 3a_1 a_{111}} + a_{11}}$.

### 3.2. Analytical and numerical solution for a homogeneous external field

Here we derive approximate analytical solution of Eqs.(11) for the case of *capacitor geometry*, $\tilde{E}_1^{ext} = \tilde{E}_2^{ext} = 0$ and small $E_3^{ext}$ under several simplifying assumptions. Using Eq.(8) for the case of semi-infinite ferroelectric (or ferroelectric film on an ideally matched substrate) yields the component of surface strains as

$$u_{i3}^S(x_1, x_3 = 0) = \frac{Q_{44}}{2} \tilde{P}_i(x_1) P_3(x_1) \quad i = 1, 2 \quad (14)$$

Unexpectedly, expressions (14) coincide with a bulk strains for arbitrary polarization profiles (see **Appendix A4** for derivation). Hence, $u_{i3}^S(\mathbf{x})/E_3^{ext}$ defines *effective piezoelectric coefficients*:

$$d_{34}^{eff} = \frac{\partial u_{23}^S(x_1, x_3 = 0)}{\partial E_3^{ext}} = \frac{Q_{44}}{2} \frac{\partial}{\partial E_3^{ext}}\left(P_3(x_1) \tilde{P}_2(x_1)\right) \quad (15a)$$

$$d_{35}^{eff} = \frac{\partial u_{13}^S(x_1, x_3 = 0)}{\partial E_3^{ext}} = \frac{Q_{44}}{2} \frac{\partial}{\partial E_3^{ext}}\left(P_3(x_1) \tilde{P}_1(x_1)\right) \quad (15b)$$

Assuming that the pinning field $E^{pin}$ is higher then the infinitely small probing field $E_3^{ext}$, one can neglect the term $O\left((E_3^{ext})^2\right)$ in expressions for $\tilde{P}_{1,2}$, since the motion of the twin wall is absent. Correspondingly approximate expressions for polarization components have the form:

$$\tilde{P}_1(\tilde{x}_1) \approx \frac{P_S}{\sqrt{2}}, \quad \tilde{P}_2(\tilde{x}_1) = \frac{P_t \sinh(\tilde{x}_1/w)}{\sqrt{A + \cosh^2(\tilde{x}_1/w)}} \approx \frac{P_S}{\sqrt{2}} \tanh\left(\frac{\tilde{x}_1}{w}\right) \qquad (16)$$

For the specific case DW should be perpendicular to $[100]$ or $[010]$ directions. Parameters:

$$P_t^2 = \frac{-a}{\sqrt{b^2 - 3ac} + b} \equiv \frac{P_S^2}{2}, \quad A = \frac{2cP_t^2}{2b + 4cP_t^2}, \quad w = \sqrt{\frac{\tilde{g}_{66}}{P_t^2(2b + 6cP_t^2)}}. \qquad (17a)$$

$$a = a_1 + \left(\frac{\tilde{a}_{12}}{2} + \frac{\tilde{a}_{112}}{4} P_S^2 - \frac{\tilde{Q}_{11}^2 s_{11} - 2Q_{12}\tilde{Q}_{11}s_{12} + Q_{12}^2 \tilde{s}_{11}}{2(s_{11}\tilde{s}_{11} - s_{12}^2)}\right) P_S^2 \qquad (17b)$$

$$b = \tilde{a}_{11} + \frac{\tilde{a}_{112}}{2} P_S^2 + \frac{\tilde{Q}_{11}^2 s_{11} - 2Q_{12}\tilde{Q}_{11}s_{12} + Q_{12}^2 \tilde{s}_{11}}{2(s_{11}\tilde{s}_{11} - s_{12}^2)}, \quad c = \tilde{a}_{111} \qquad (17c)$$

Parameter $w$ is the width of the domain wall affected by electrostriction coupling. Note, that solution (16) is the solution of equation $2aP_2 + 4bP_2^3 + 6cP_2^5 - g\frac{\partial^2 P_2}{\partial \tilde{x}_1^2} = 0$ (see **Appendix A3**).

Simple analytical expression for $P_3(\tilde{x}_1)$ is valid for small gradient terms and linearization with respect to $P_3$ (see **Appendix A4**):

$$P_3(\tilde{x}_1) \approx E_3^{ext} \chi_{33}^{eff}(\tilde{x}_1), \qquad (18a)$$

$$\chi_{33}^{eff} = \left(2\left(a_1 + a_{12}\left(\frac{P_S^2}{2} + \tilde{P}_2^2\right) + \tilde{a}_{113}\left(\frac{P_S^4}{4} + \tilde{P}_2^4\right) + \tilde{a}_{123}\frac{P_S^2}{2}\tilde{P}_2^2 - Q\left(\frac{P_S^2}{2} - \tilde{P}_2^2\right)\right) + \frac{Q_{44}^2 \tilde{P}_2^2}{s_{44}} + \frac{g_{44}}{w^2}\right)^{-1} \qquad (18b)$$

Where parameter $Q = Q_{12}\frac{s_{11}\tilde{Q}_{11} - s_{12}Q_{12}}{s_{11}\tilde{s}_{11} - s_{12}^2} + Q_{11}\frac{\tilde{s}_{11}Q_{12} - s_{12}\tilde{Q}_{11}}{s_{11}\tilde{s}_{11} - s_{12}^2}$. Note, that the function $\chi_{i3}^{eff}$ can be treated as the effective susceptibility. Equations (18) are exact under the condition $\tilde{g}_{ij} \to 0$; it can be rather accurate positive and almost constant denominator, that may appear not the case in the vicinity of TW, where the gradient effect should be accounted more accurately.

**Figures 2-3** show the $x_1$-profile of the effective piezoelectric coefficients $d_{34}^{eff} = \partial u_{23}^S / \partial E_3^{ext}$, $d_{35}^{eff} = \partial u_{13}^S / \partial E_3^{ext}$. Note that piezoelectric response of $a_1$-$a_2$ twins is practically independent on the flexoelectric coupling, which have impact only on the polarization component $\tilde{P}_1$.

**Figure 2** presents dependence of polarization components $\tilde{P}_i$ **(a, b)** on the distance $\tilde{x}_1$ from the wall in 45°-rotated coordinate system. Note that the $\tilde{P}_3$ component is independent on distance $\tilde{x}_1$; it was set at zero in calculations. The $\tilde{P}_1$ component is the even function of the $\tilde{x}_1$ - coordinate, that very weakly depends on the distance $\tilde{x}_1$, so it was set as constant in the calculations. In more detail the spatial region where the component $\tilde{P}_1$ varies is scaled in the plot **(b)**, where it is calculated for different flexoelectric coefficients. The $\widetilde{P}_1$ component has maximum at $\widetilde{x}_1 \approx 0$, which position is weakly dependent on the flexoelectric coupling value, then the component decreases to a constant value with the distance from the wall. The $\widetilde{P}_2$ component is the odd function of the $\widetilde{x}_1$ - coordinate, that saturates to a constant values on the long distance from the wall. It is worth noting that on the long distance from the wall both $\widetilde{P}_1$ and $\widetilde{P}_2$ components have the same absolute value.

Plots **(c,d)** show the $x_1$-profile of the effective piezoelectric coefficients $d_{34}^{eff}$ and $d_{35}^{eff}$ calculated across a twin wall in tetragonal phase of BaTiO$_3$ at room temperature for electric field $E_3^{ext} \to 0$. Plot **(d)** is calculated numerically. Plot **(d)** comprises the result of analytic calculation of the same coefficient with and without taking into account the polarization gradient term. Including the gradient term leads to more accurate convergence with numerical estimation for $d_{35}^{eff}$. However, this trend is absent for the $d_{34}^{eff}$ coefficient, because in this case the presence of the gradient term decreases the magnitude of local maximum (minimum) and slightly moves it away from the wall. In more detail the influence of the gradient term on the $d_{34}^{eff}$ and $d_{35}^{eff}$ $x_1$-profiles is illustrated on the **Figure 3**. The dependences shown in this figure were calculated for different values of gradient coefficients: 0.1 $g_{44}$, $g_{44}$ and 10 $g_{44}$ respectively. The $d_{35}^{eff}$ calculated for 10 $g_{44}$ (curve 3) is about two times of magnitude smaller than $d_{35}^{eff}$ calculated for 0.1 $g_{44}$ (curve 1); also curve 3 is more narrow than the curve 1. For $d_{34}^{eff}$ coefficient, as it was mentioned above, increasing of the gradient term value leads to the slight decrease of the magnitude of local maximum (minimum). Local maximum (minimum) also moved away from the wall and become more extended with increasing of gradient term value. Plot **(c)** presents the effective susceptibility tensor component $\chi_{33}^{eff}$, and the arguments given for the $d_{35}^{eff}$ curve are also applicable there. $X_1$-profile of potential barrier across the TW is shown in **Fig. 3d** for different flexoelectric coefficients. One can see that the barrier is weakly affected by the flexoelectric coupling strength.

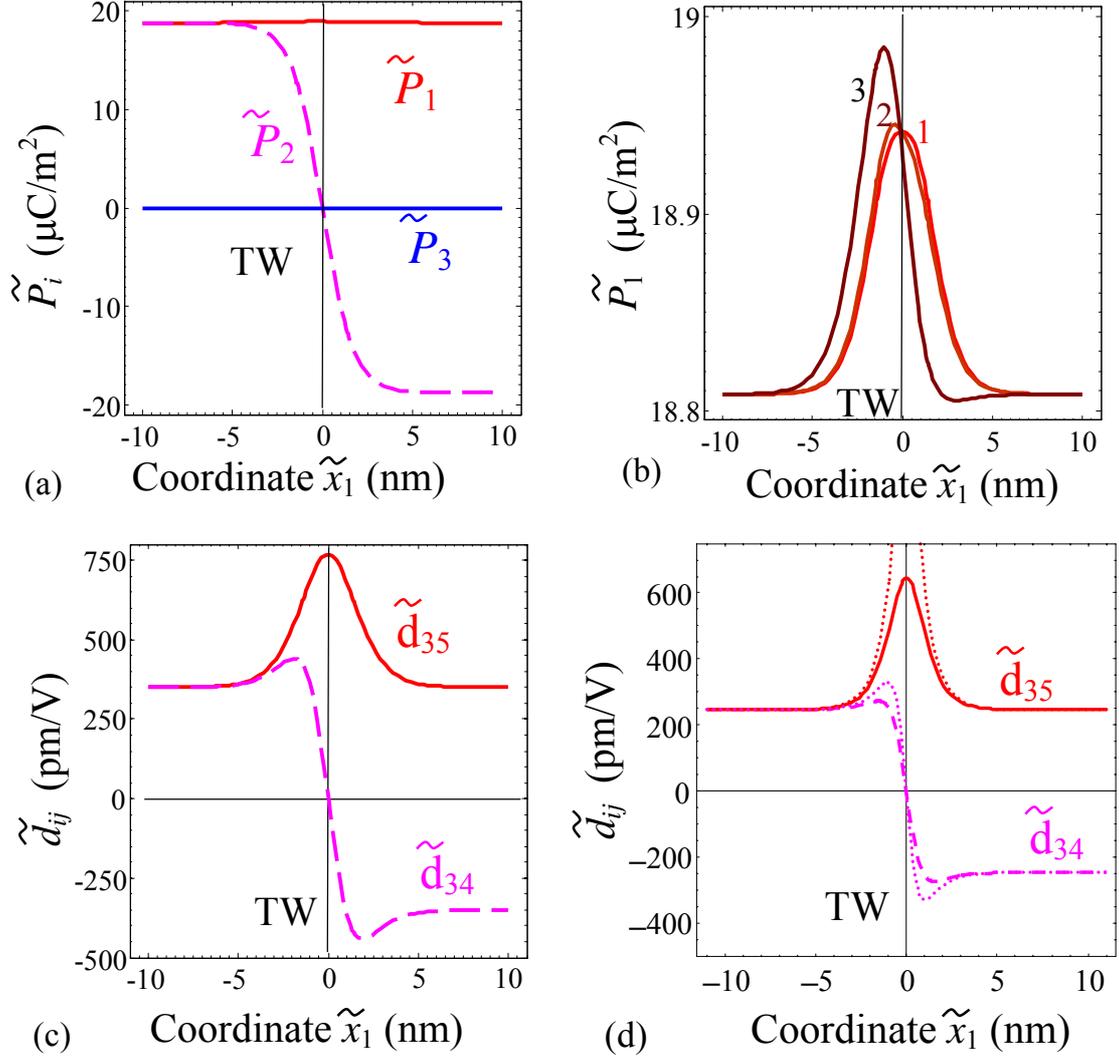

**FIGURE 2.** X$_1$-profile of the polarization components **(a,b)**, effective piezoelectric coefficient $d_{34}^{eff}$ and $d_{35}^{eff}$ **(c,d)** calculated across a twin wall in tetragonal phase of BaTiO$_3$ at room temperature for electric field $E_3^{ext} \to 0$. Coefficients **(c)** were calculated numerically from coupled Eq.(11) at $\tilde{g}_{ij} \neq 0$; **(d)** were calculated analytically from Eqs.(18) at $\tilde{g}_{ij} = 0$ (dotted curves). Profiles **(b)** of $\tilde{P}_1(\tilde{x}_1)$ are calculated for different flexoelectric coefficients 0, $f_{ij}$, 2$f_{ij}$, (curves 1, 2, 3 respectively), where $f_{ij}$ along with other parameters and LGD expansion coefficients are listed in the **Table A3.**

Finally, we analyze the situation when a twin wall can freely move in the ***bulk*** even in a very small external field (no pinning). For the case effective piezoelectric constants were calculated as $d_{ijk}^{eff} = \partial u_{jk}^S / \partial E_i^{ext}$ assuming that all components of external field are infinitely small, $E_i^{ext} \to 0$. Results are shown in the **Figure 4.** Comparison of the figures 2, 3 and 4 illustrates the pronounced differences in the piezoelectric response of the "free" twin walls in the bulk without pinning and the wall-surface junction with a lattice pinning. In particular effective piezoelectric response across the twin wall can be much higher than in the single-domain region. In numbers, the enhancement reaches $10^3$ times for $d_{26}^{eff}$, without consideration of pinning effects.

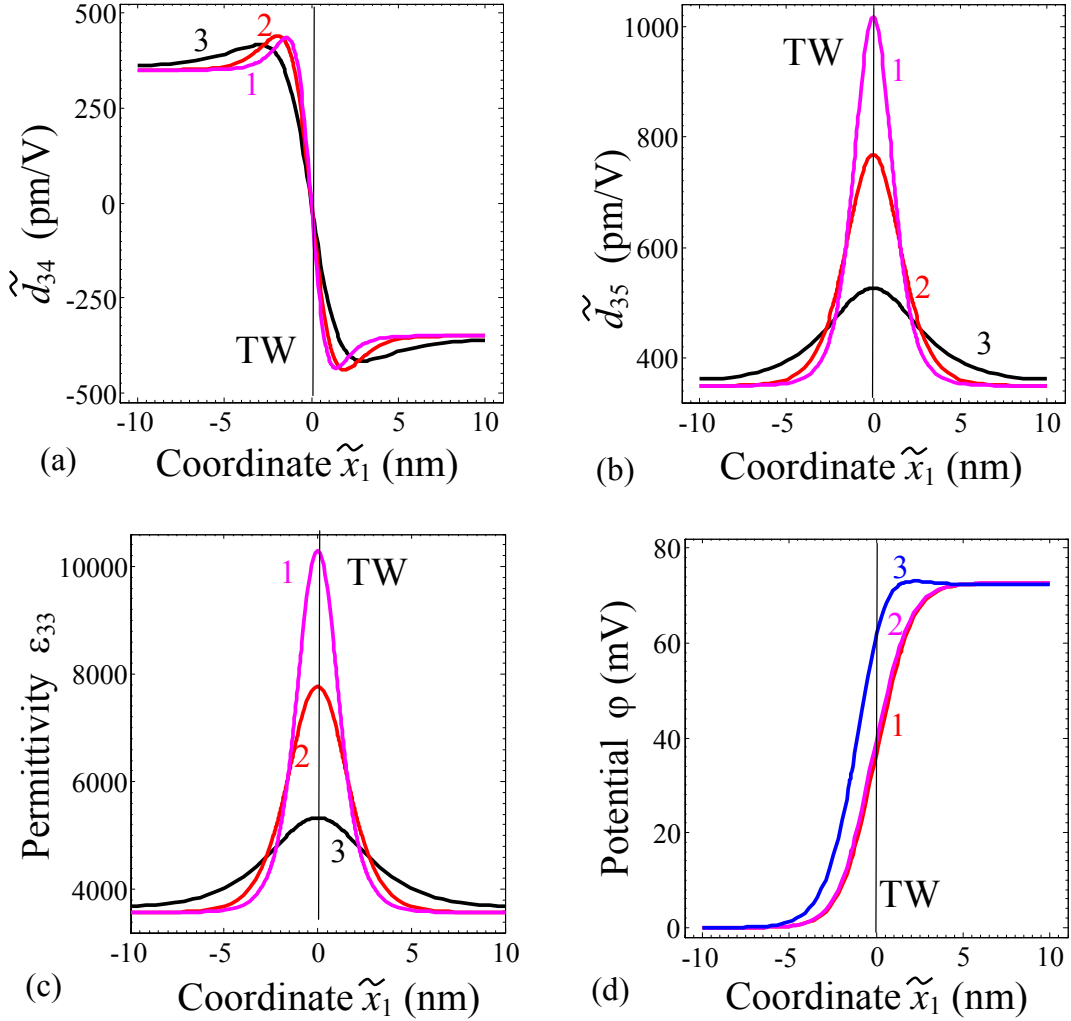

**FIGURE 3.** $X_1$-profile of the effective piezoelectric coefficients $d_{34}^{eff}$ **(a)**, $d_{35}^{eff}$ **(b)** and susceptibility $\chi_{33}^{eff}$ **(c)** calculated numerically from coupled Eq.(11) across a twin wall in tetragonal phase of BaTiO$_3$ at room temperature, zero flexoelectric coefficients and different gradient coefficients 0.1 $g_{44}$, $g_{44}$, 10 $g_{44}$, (curves 1, 2, 3 respectively). **(d)** $x_1$-profile of potential barrier for different flexoelectric coefficients 0 $f_{ij}$, $f_{ij}$, 2 $f_{ij}$, (curves 1, 2, 3 respectively) and room temperature. Electric field $E_3^{ext} \to 0$. Coefficients $g_{ij}$ and $f_{ij}$ along with other parameters are listed in the Table A3.

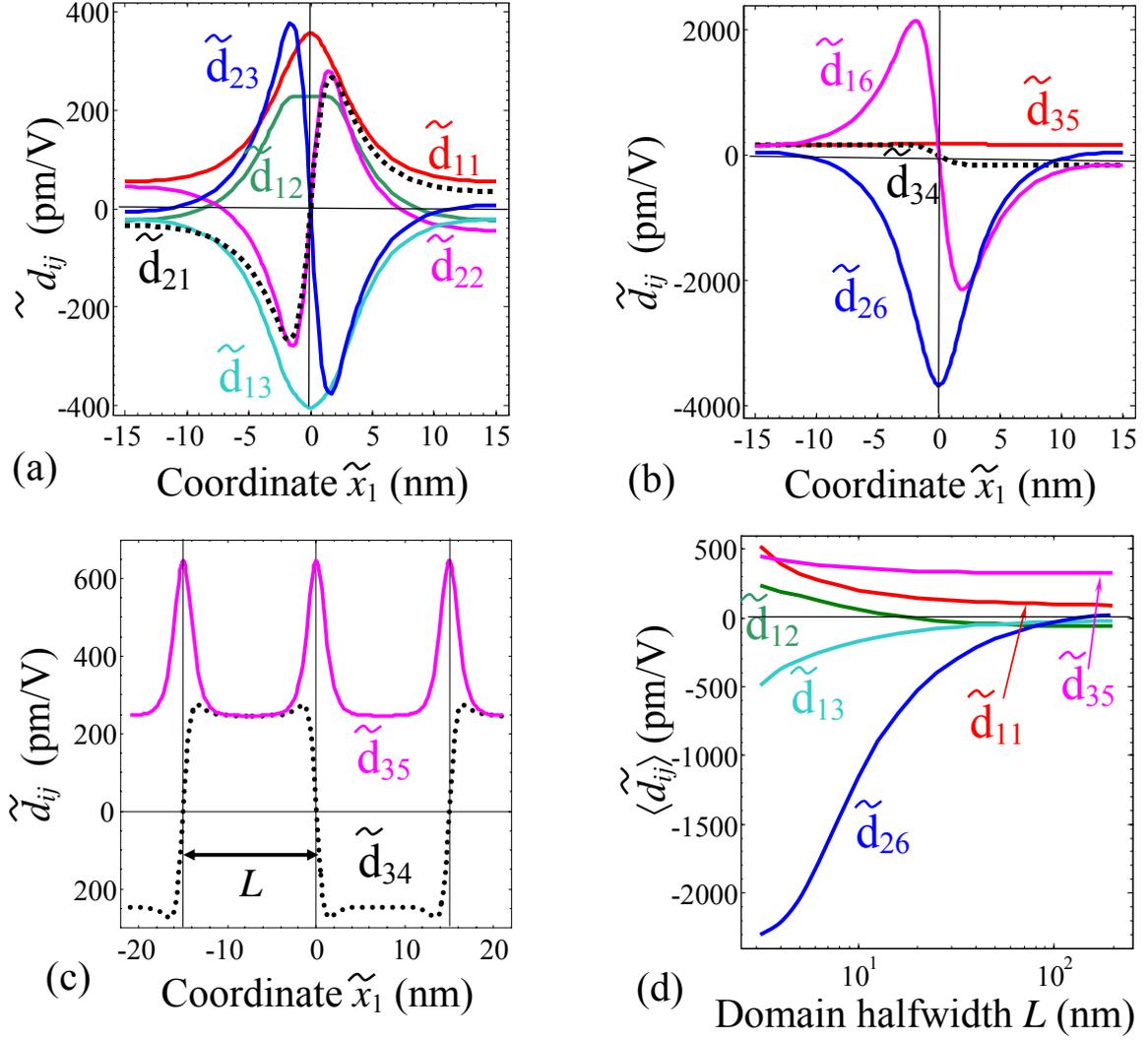

**FIGURE 4. Effective piezoelectric coefficients without pinning.** Distribution of dilatational **(a)** and shear **(b)** components of piezoelectric tensor $\tilde{d}_{ijk}$ across {110} twin wall in tetragonal phase of bulk BaTiO$_3$ at room temperature. Twin wall (TW) boundary is $\tilde{x}_1 = 0$. **(c)** Profiles and **(d)** average values of piezoelectric coefficients of the periodic twins. LGD expansion coefficients are summarized in **Table A3**.

These predictions can be directly probed by PFM observations of ferroelectric domain walls in electroded thinned single crystals (but note the effect of mechanical coupling on PFM resolution [22]). However, it is also of interest to explore the effects on macroscopic properties. In the case when we deal with a dense lamellar $a_1$-$a_2$ domain structures of half-period $L$ (as shown in **Figure 4c**), average values of the even-type components of the piezoelectric response $\langle \tilde{d}_{11} \rangle$, $\langle \tilde{d}_{12} \rangle$, $\langle \tilde{d}_{13} \rangle$, $\langle \tilde{d}_{26} \rangle$ and $\langle \tilde{d}_{35} \rangle$ become stronger (**Figure 4d**); odd-type components $\langle \tilde{d}_{21} \rangle$, $\langle \tilde{d}_{22} \rangle$, $\langle \tilde{d}_{23} \rangle$, $\langle \tilde{d}_{16} \rangle$ and $\langle \tilde{d}_{34} \rangle$ vanish. One can see from the **Figure 4d** that nonzero $\langle \tilde{d}_{ij} \rangle$ can be much higher for dense lamellar domain of period $L$=10 nm; it monotonically decreases with the twin walls separation increase and reaches bulk values for $L$>100 nm.

## 4. SUMMARY

Here, we explored electromechanical phenomena on the twin domain walls in tetragonal ferroelectrics as relevant to PFM studies and domain-wall controlled materials functionalities. Piezoelectric response of *aa* twins is practically independent on the flexoelectric coupling, but extremely sensitive to the values of polarization gradient coefficients, since the local dielectric susceptibility is very sensitive to the gradient coefficients. The difference between the surface strains and "unperturbed" bulk ones does not exist in the first order of perturbation theory, but may be essential in higher orders.

The effective piezoelectric response across the twin wall can be (much) higher than in the single-domain region. The enhancement ranging from 10% for $d_{34}^{eff}$, 200% for $d_{35}^{eff}$ for pinned walls and up to the giant values, like $10^3$ times for $d_{26}^{eff}$, without consideration of pinning effects. Considering a dense lamellar $a_1$-$a_2$ domain structures of half-period $L$, average values of the odd-type components of the piezoelectric response vanish, while the even-type components become can be much higher for dense lamellar domain of period $L$=10 nm; they monotonically decreases with the twin walls separation increase and reaches bulk values for $L$>100 nm.

The effective piezoelectric response is proportional to the local dielectric susceptibility, i.e. if $d_{26}^{eff}$ increases in n-times across the twin wall, the same n-times enhancement corresponds to the local susceptibility, as anticipated from the linear mechanism of the piezoelectric response formation. We did not find any significant deviation from the proportionality law and therefore the intriguing case, when piezoelectric response is high, but the dielectric susceptibility enhancement is absent or moderate, was not reached.

**Acknowledgements**
Authors gratefully acknowledge multiple discussions, critical remarks and useful suggestions about local piezoelectric response and dielectric permittivity calculations from Prof. Alexander K. Tagantsev (EPFL, Lausanne, Switzerland). The work was supported in part (SVK) by the office Basic Energy Sciences, US Department of Energy. A.N.M. and E.A.E. acknowledge State Fund of Fundamental State Fund of Fundamental Research of Ukraine, SFFR-NSF project UU48/002 (NSF number DMR-1210588).

## Supplemental Materials
## Appendix A

**A1. Free energy**

In order to model the electrical and elastic properties of ionic semiconductor in equilibrium, we derive the generalized expression for the free energy functional. Free energy for cubic symmetry paraelectrics including quantum corrections has the following form:

$$F = F_{ES} + F_{FLEXO} + F_{SE} + F_{CS} \tag{A.1a}$$

The first term in Eq.(A.1a) is the electrostatic energy of the quantum paraelectric with soft phonon mode, polarization gradient and space charge, that can be written down as

$$F_{ES} = \int_V d^3r \left( \frac{\alpha(T)}{2} P_i P_i + \alpha_{ijkl} P_i P_j P_k P_l + \frac{g_{ijkl}}{2} \left( \frac{\partial P_i}{\partial x_j} \frac{\partial P_k}{\partial x_l} \right) - P_i E_i - \frac{\varepsilon_0 \varepsilon_b}{2} E_i E_i \right) \tag{A.1b}$$

Hereinafter $P_m(\mathbf{r})$ denotes electric polarization.

The second term in Eq.(1a) is the flexoelectric effect contribution:

$$F_{FLEXO} = \int_V d^3r \frac{f_{ijkm}}{2} \left( u_{ij} \frac{\partial P_m}{\partial x_k} - P_m \frac{\partial u_{ij}}{\partial x_k} \right). \tag{A.1c}$$

Hereinafter $u_{kl}(\mathbf{r})$ denotes the elastic strain and $f_{ijkm}$ the flexoelectric tensor.

$$F_{SE} = \int_V d^3r \left( u_{ij} q_{ijkl} P_k P_l + \frac{c_{ijkl}}{2} u_{ij} u_{kl} \right) \tag{A.1d}$$

Here $q_{ijkl}$ denotes the electrostriction stress tensor, $c_{ijkl}$ the elastic stiffness tensor.

**A2. Green function**

$$G_{ij}(x_1, x_2, \xi_3) = \begin{cases} \frac{1+\nu}{2\pi Y} \left[ \frac{\delta_{ij}}{R} + \frac{(x_i - \xi_i)(x_j - \xi_j)}{R^3} + \frac{1-2\nu}{R+\xi_3} \left( \delta_{ij} - \frac{(x_i - \xi_i)(x_j - \xi_j)}{R(R+\xi_3)} \right) \right] & i,j \neq 3 \\ \frac{(1+\nu)(x_i - \xi_i)}{2\pi Y} \left( \frac{-\xi_3}{R^3} - \frac{(1-2\nu)}{R(R+\xi_3)} \right) & i = 1,2 \text{ and } j = 3 \\ \frac{(1+\nu)(x_j - \xi_j)}{2\pi Y} \left( \frac{-\xi_3}{R^3} + \frac{(1-2\nu)}{R(R+\xi_3)} \right) & j = 1,2 \text{ and } i = 3 \\ \frac{1+\nu}{2\pi Y} \left( \frac{2(1-\nu)}{R} + \frac{\xi_3^2}{R^3} \right) & i = j = 3 \end{cases} \tag{A.2}$$

Here $R = \sqrt{(x_1 - \xi_1)^2 + (x_2 - \xi_2)^2 + \xi_3^2}$ is radius vector, $Y$ is Young's modulus, and $\nu$ is the Poisson ratio. Stiffness tensor $c_{kjmn}$ corresponds to the elastically isotropic medium

$$c_{klmn} = \frac{Y}{2(1+\nu)}\left[\frac{2\nu}{1-2\nu}\delta_{kl}\delta_{mn} + \delta_{km}\delta_{ln} + \delta_{kn}\delta_{lm}\right].$$

**Appendix A3.**

Euler-Lagrange equation

$$2aP_2 + 4bP_2^3 + 6cP_2^5 - g\frac{\partial^2 P_2}{\partial \tilde{x}_1^2} = 0 \qquad (A.3)$$

Has a well-known solution:

$$P_2(\tilde{x}_1) = \frac{P_t \sinh(\tilde{x}_1/w)}{\sqrt{A + \cosh^2(\tilde{x}_1/w)}} \qquad (A.4a)$$

$$P_t^2 = \frac{-a}{\sqrt{b^2 - 3ac} + b}, \quad w = \sqrt{\frac{g}{P_t^2(2b + 6cP_t^2)}}, \quad A = \frac{2cP_t^2}{2b + 4cP_t^2} \qquad (A.4b)$$

**Table A3.** Free energy coefficients for bulk ferroelectric $BaTiO_3$ (from Refs.[46, 47]).

| parameters | values | Refs. and Notes |
|---|---|---|
| $a_1$ ($C^{-2}$·mJ) | $a_1 = 3.34(T-381)\times 10^5$ | Ref. [a] |
| $a_{ij}$ ($C^{-4}$·m$^5$J) | $a_{11} = 4.69(T-393)\times 10^6 - 2.02\times 10^8$ <br> $a_{12} = 3.230\times 10^8$ | Ref. [a] |
| $a_{ijk}$ ($C^{-6}$·m$^9$J) | $a_{111} = -5.52(T-393)\times 10^7 + 2.76\times 10^9$ <br> $a_{112} = 4.47\times 10^9$ <br> $a_{123} = 4.91\times 10^9$ | Ref. [a] |
| $g_{ij}$ ($C^{-2}$m$^3$J) | $g_{11} = 5.1\times 10^{-10}$ <br> $g_{12} = -0.2\times 10^{-10}$ <br> $g_{44} = 0.2\times 10^{-10}$ | Ref. [b] |
| $Q_{ij}$ ($C^{-2}$·m$^4$) | $Q_{11} = 0.11$, $Q_{12} = -0.045$, $Q_{44} = 0.029$ | Recalculated from Ref.[b] <br> Different from Ref. [c] |
| $s_{ij}$ ($10^{-12}$ Pa$^{-1}$) | $s_{11} = 7.47$, $s_{12} = -2.95$, $s_{44} = 18.42$ | Recalculated from Ref.[b] <br> Different from Ref. [c] |
| $f_{ijkl}$ (V) | $f_{11} = 5.12$, $f_{12} = 3.32$, $f_{44} = 0.045$; [48] | Ref. [d] |
| $F_{ijkl}$ ($10^{-11}$C$^{-1}$m$^3$) | $F_{11} = 2.46$, $F_{12} = 0.48$, $F_{44} = 0.05$ | recalculated using $F_{\alpha\gamma} = f_{\alpha\beta}s_{\beta\gamma}$ |

[a] A.J. Bell. J. Appl. Phys. **89**, 3907 (2001).
[b] P. Marton, I. Rychetsky, and J. Hlinka. Phys. Rev. B **81**, 144125 (2010).
[c] N.A. Pertsev, A. G. Zembilgotov, and A. K. Tagantsev. Phys. Rev. Lett. 80, 1988 (1999).

[d] I. Ponomareva, A. K. Tagantsev, L. Bellaiche. Phys. Rev. B 85, 104101 (2012).

**Appendix A4.**

Using Eq.(4) we could get the only component of strains depending on $P_3$ as

$$u_{23}^S(\mathbf{x}) = \frac{Q_{44}}{2} \int_{-\infty}^{\infty} \frac{x_3 \tilde{P}_2(\xi_1) P_3(\xi_1) d\xi_1}{\pi\left(x_3^2 + (x_1 - \xi_1)^2\right)} \tag{A.5}$$

Here we used $s_{44} = 2(1+\nu)/Y$. Note that the integral in (A.4) could be obtained exactly for arbitrary polarization profile at the **surface**, i.e. at $x_3 \to 0$, since in this case the convolution kernel is the one of δ-function representations:

$$u_{23}^S(x_1, x_3 = 0) = \frac{Q_{44}}{2} \tilde{P}_2(x_1) P_3(x_1) \tag{A.6}$$

Using slanted step approximation for $\tilde{P}_2(x_1)$, one could get the following expression for $u_4^S(\mathbf{x}) = 2u_{23}^S(\mathbf{x})$:

$$u_4^S \approx P_3(x_1) \frac{Q_{44} P_S}{\pi\sqrt{2}} \left( \begin{array}{c} \left(1 - \frac{x_1}{w}\right) \arctan\left(\frac{w - x_1}{x_3}\right) - \left(1 + \frac{x_1}{w}\right) \arctan\left(\frac{w + x_1}{x_3}\right) \\ + \frac{x_3}{a} \ln\left(\sqrt{\frac{(w + x_1)^2 + x_3^2}{(w - x_1)^2 + x_3^2}}\right) \end{array} \right) \tag{A.7a}$$

$$d_{34}^{eff} \approx \frac{Q_{44} P_S}{\pi\sqrt{2}} \left( \begin{array}{c} \left(1 - \frac{x_1}{w}\right) \arctan\left(\frac{w - x_1}{x_3}\right) - \left(1 + \frac{x_1}{w}\right) \arctan\left(\frac{w + x_1}{x_3}\right) \\ + \frac{x_3}{a} \ln\left(\sqrt{\frac{(w + x_1)^2 + x_3^2}{(w - x_1)^2 + x_3^2}}\right) \end{array} \right) \left.\frac{\partial(P_3(x_1))}{\partial E_3^{ext}}\right|_{E_3^{ext}=0} \tag{A.7b}$$

There is another coefficient, $d_{35}^{eff}$, related with response for $E_3^{ext}$, but it is constant for our approximations.

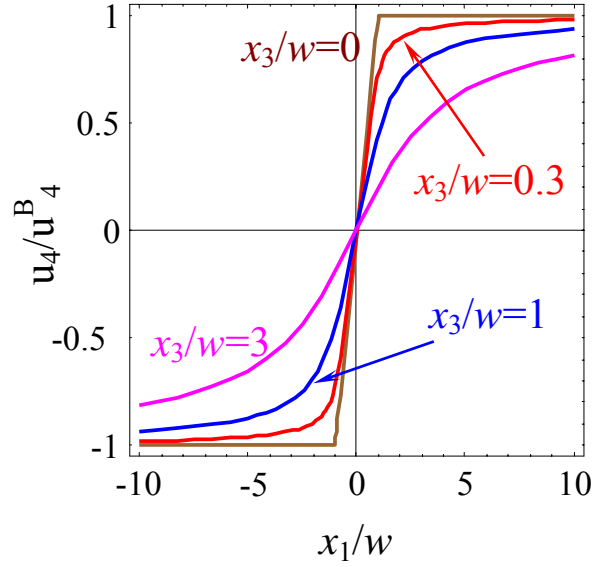

**FIGURE A1.** $X_1$-profile of the of shear strain component near the twin domain wall junction with free surface for different distance from the surface (depth), specified near the curves. LGD expansion coefficients are summarized in **Table A3.**

**Note.** Equation (18b) can be derived from the expression:

$$\chi_{33}^{eff}(\tilde{x}_1) \approx \left( 2 \begin{pmatrix} a_1 + a_{12}(\tilde{P}_1^2 + \tilde{P}_2^2) + \tilde{a}_{113}(\tilde{P}_1^4 + \tilde{P}_2^4) \\ + \tilde{a}_{123}\tilde{P}_1^2\tilde{P}_2^2 - Q_{12}\tilde{\sigma}_{22}^0 - Q_{11}\tilde{\sigma}_{33}^0 \end{pmatrix} + \frac{Q_{44}^2\tilde{P}_2^2}{s_{44}} + \frac{g_{44}}{w^2} \right)^{-1} \quad (A.8)$$

**References**

1 M. Daraktchiev, G. Catalan, J. F. Scott, Ferroelectrics, 375, 122 (2008).

[2] A.P. Pyatakov and A.K. Zvezdin, Eur. Phys. J. B 71, 419–427 (2009).

[3] J. Seidel, L.W. Martin, Q. He, Q. Zhan, Y.-H. Chu, A. Rother, M. E. Hawkridge, P. Maksymovych, P. Yu, M. Gajek, N. Balke, S. V. Kalinin, S. Gemming, F. Wang, G. Catalan, J. F. Scott, N. A. Spaldin, J. Orenstein and R. Ramesh. Nature Materials 8, 229 – 234 (2009)

[4] J. Seidel, P. Maksymovych, Y. Batra, A. Katan, S.-Y. Yang, Q. He, A. P. Baddorf, S.V. Kalinin, C.-H. Yang, J.-C. Yang, Y.-H. Chu, E. K. H. Salje, H. Wormeester, M. Salmeron, and R. Ramesh, Phys. Rev. Lett. 105, 197603 (2010)

[5] R. K. Vasudevan, A. N. Morozovska, E. A. Eliseev, J. Britson, J.-C. Yang, Y.-H. Chu, P. Maksymovych, L. Q. Chen, V. Nagarajan, S. V. Kalinin. Nano Letters, **12** (11), pp 5524–5531 **(**2012)

[6] A. Gruverman, A. Kholkin. Rep. Prog. Phys. 69, 2443–2474. (2006).

[7] S.V. Kalinin, A.N. Morozovska, Long Qing Chen, Brian J. Rodriguez. Rep. Prog. Phys. 73, 056502-1-67 (2010).

[8] A. Gruverman, Ferroelectrics, 433, 88-106, (2012).

[9]. A. Gruverman, O. Auciello, H. Tokumoto. *Annu. Rev. Mater. Sci.* **28,** 101 (1998).

[10] A. Wu, P.M. Vilarinho, V.V. Shvartsman, G. Suchaneck, A.L. Kholkin, Nanotechnology, 16, 2587 (2005).

11. V.V. Shvartsman, N.A. Pertsev, J.M. Herrero, C. Zaldo, and A.L. Kholkin *J. Appl. Phys. 97,* 104105 (2005).

[12]. V.V. Shvartsman, A.L. Kholkin, M. Tyunina and J. Levoska. *Appl. Phys. Lett.* **86** 222907 (2005).

[13] S.V. Kalinin, B.J. Rodriguez, S. Jesse, P. Maksymovych, K. Seal, M. Nikiforov, A.P. baddorf, A.L. Kholkin, and R. Proksch, Materials Today **11**, 16 (2008).

[14]. N.A. Pertsev, A. Petraru, H. Kohlstedt, R. Waser, I.K. Bdikin, D. Kiselev and A.L. Kholkin, *Nanotechnology* **19**, 375703 (2008).

[15] C.S. Ganpule, V. Nagarjan, H. Li, A.S. Ogale, D.E. Steinhauer, S. Aggarwal, E. Williams, R. Ramesh and P. De Wolf *Appl. Phys. Lett.* **77** 292 (2000).

[16] A. Agronin, M. Molotskii, Y. Rosenwaks, E. Strassburg, A. Boag, S. Mutchnik and G. Rosenman. *J. Appl. Phys.* **97** 084312 (2005).

[17] F. Felten, G.A. Schneider, J. Muñoz Saldaña, S.V. Kalinin. *J. Appl. Phys.* **96** 563 (2004).

[18] D.A. Scrymgeour and V. Gopalan. *Phys. Rev. B* **72** 024103 (2005).

[19] S.V. Kalinin, E.A. Eliseev and A.N. Morozovska *Appl. Phys. Lett.* **88** 232904 (2006).

[20] E.A. Eliseev, S.V. Kalinin, S. Jesse, S.L. Bravina, A.N. Morozovska. *J. Appl. Phys.* **102** 014109 (2007).

[21] A.N. Morozovska, E.A. Eliseev, S.L. Bravina, S.V. Kalinin *Phys. Rev. B* **75** 174109 (2007).

[22] S.V. Kalinin, B.J. Rodriguez, S.-H. Kim, S.-K. Hong, A. Gruverman, E.A. Eliseev. Appl. Phys. Lett. 92, № 15, 152906 (2008).

[23] Shiming Lei, Eugene A. Eliseev, Anna N. Morozovska, Ryan C. Haislmaier, Tom T. A. Lummen, W. Cao, Sergei V. Kalinin and Venkatraman Gopalan. Phys. Rev. B 86, 134115 (2012).

[24] K. Pan, Y. Y. Liu, Y. M. Liu, and J. Y. Li: J. Appl. Phys. 112, 052016. (2012)

[25] K. Pan, Y. M. Liu, Y. Y. Liu, and J. Y. Li. Resolving ferroelectric nanostructures via piezoresponse force microscopy - a numerical investigation. Applied Physics Express (2012)

[26] Lun Yang and Kaushik Dayal. Modelling Simul. Mater. Sci. Eng. 20, 035021 (2012).

[27] Lun Yang and Kaushik Dayal. J. Appl. Phys. 111, 014106 (2012);

[28] Tomas Sluka, Alexander K. Tagantsev, Dragan Damjanovic, Maxim Gureev, Nava Setter. Nature Com. 3, 748 (2012)

[29] J. Karthik, A. R. Damodaran, and L. W. Martin. Phys. Rev. Lett. 108, 167601 (2012).

[30] A.N. Morozovska, S.V. Kalinin, E.A. Eliseev, V. Gopalan, S.V. Svechnikov. Phys. Rev. B. 78,

125407 (2008).

[31] Vasudeva Rao Aravind, A.N. Morozovska, S. Bhattacharyya, D. Lee, S. Jesse, I. Grinberg, Y.L. Li, S. Choudhury, P. Wu, K. Seal, A.M. Rappe, S.V. Svechnikov, E.A. Eliseev, S.R. Phillpot, L.Q. Chen, Venkatraman Gopalan, S.V. Kalinin. Phys. Rev. B **82**, 024111-1-11 (2010).

[32] G.A. Korn, and T.M. Korn. Mathematical handbook for scientists and engineers (McGraw-Hill, New-York, 1961).

[33] N.A. Pertsev, A.G. Zembilgotov, and A.K. Tagantsev, Phys. Rev. Lett. **80**, 1988 (1998).

[34] E.A. Eliseev and A.N. Morozovska, J. Mater. Sci. **44**, 5149 (2009).

[35] K. Schwarz, U. Rabe, S. Hirsekorn, and W. Arnold, Appl. Phys. Lett. **92**, 183105 (2008).

[36] S. M. Sze, *Physics of Semiconductor Devices*, 2nd ed. Ch.5 (Wiley-Interscience, New York, 1981).

[37] S.P. Timoshenko and J.N. Goodier, *Theory of Elasticity*, McGraw-Hill, N. Y., 1970.

[38] I Rychetsky, J. Phys.: Condens. Matter 9, 4583–4592 (1997).

[39] A.I. Lur'e, *Three-dimensional problems of the theory of elasticity* (Interscience Publishers, 1964).

[40] L.D. Landau and E.M. Lifshitz, *Theory of Elasticity*. Theoretical Physics, Vol. 7 (Butterworth-Heinemann, Oxford, U.K., 1998).

[41] W. Cao, .E. Cross, Phys.Rev. B 44, 5 (1991)

[42] Supplemental Materials to Anisotropic conductivity of uncharged domain walls in $BiFeO_3$. *Accepted to Phys. Rev. B* (http://arxiv.org/abs/1206.1289) by Anna N. Morozovska, Rama K. Vasudevan, Peter Maksymovych, Sergei V. Kalinin and Eugene A. Eliseev.

[43] Y. Ishibashi, J. Phys. Soc. Jpn. **46**, 1254 (1979).

[44] I. Suzuki and Y. Ishibashi, Ferroelectrics, **64**, 181 (1985).

[45] S. Choudhury, Y. Li, N. Odagawa, Aravind Vasudevarao, L. Tian, P. Capek, V. Dierolf, A. N. Morozovska, E.A. Eliseev, S.V. Kalinin, Y. Cho, L.-Q. Chen, V. Gopalan. J. Appl. Phys. 104, № 8, 084107 (2008).

[46] A.J. Bell. J. Appl. Phys. **89**, 3907 (2001).

[47] P. Marton, I. Rychetsky, and J. Hlinka. Phys. Rev. B **81**, 144125 (2010).

[48] I. Ponomareva, A. K. Tagantsev, L. Bellaiche. Phys. Rev. B 85, 104101 (2012).